\documentclass[a4]{article}
\usepackage[latin1]{inputenc}
\usepackage[dvips]{graphicx,epsfig,color}
\usepackage{wrapfig,rotating}
\usepackage{amssymb,amsmath,array}

\newcommand{\sub}[1]{\ensuremath{_\mathrm{#1}}}
\newcommand{\super}[1]{\ensuremath{^\mathrm{#1}}}
\newcommand{\MeV}{\ensuremath{\mathrm{MeV}}}
\newcommand{\GeV}{\ensuremath{\mathrm{GeV}}}
\newcommand{\pb}{\ensuremath{\mathrm{pb}}}
\newcommand{\fb}{\ensuremath{\mathrm{fb}}}
\newcommand{\ps}{\ensuremath{\mathrm{ps}}}
\newcommand{\lumi}{\ensuremath{\mathcal{L}}}
\newcommand{\percmsqs}{\ensuremath{\mathrm{cm^{-2}\,s^{-1}}}}
\newcommand{\pt}{\ensuremath{p\sub t}}

\begin{document}
\title{Summary of the Heavy Flavor Working Group
\footnote{To apper in the Proceedings of the XV International Workshop 
on
Deep-Inelastic Scattering and Related Subjectes, DIS 2007, April 16--20, 2007, Munich, Germany.}
}

\author{Michael Klasen$^1$, Benno List$^2$, \\
Stephanie Hansmann-Menzemer$^3$, 
 and Rainer Mankel$^4$
%
%
\vspace{.3cm}\\
%
1- Universit\'e Grenoble I - 
Laboratoire de Physique Subatomique et de
 Cosmologie \\
 53 Avenue des Martyrs - 38026 Grenoble - France
%
\vspace{.1cm}\\
2- Universit\"at Hamburg - Institut f\"ur Experimentalphysik \\
Luruper Chaussee 149 - 22761 Hamburg - Germany
\vspace{.1cm}\\
3- Universit\"at Heidelberg - Physikalisches Institut \\
Philosophenweg 12 - 69120 Heidelberg - Germany 
\vspace{.1cm}\\
4- Deutsches Elektronen-Synchrotron DESY - F1 \\
Notkestr. 85 - 22603 Hamburg - Hamburg
}

\begin{titlepage}

\noindent
\begin{flushright} 
LPSC 07-73
\end{flushright}

\vspace{2cm}
\begin{center}
\begin{LARGE}

{\bf Summary of the Heavy Flavor Working Group}
\end{LARGE}

\vspace{2cm}

Michael Klasen$^1$, Benno List$^2$, 
Stephanie Hansmann-Menzemer$^3$, 
 and Rainer Mankel$^4$
%
%
{\footnotesize
\vspace{.3cm}\\
%
1- Universit\'e Grenoble I - \\
Laboratoire de Physique Subatomique et de
 Cosmologie \\
 53 Avenue des Martyrs - 38026 Grenoble - France
%
\vspace{.1cm}\\
2- Universit\"at Hamburg - Institut f\"ur Experimentalphysik \\
Luruper Chaussee 149 - 22761 Hamburg - Germany
\vspace{.1cm}\\
3- Universit\"at Heidelberg - Physikalisches Institut \\
Philosophenweg 12 - 69120 Heidelberg - Germany 
\vspace{.1cm}\\
4- Deutsches Elektronen-Synchrotron DESY - F1 \\
Notkestr. 85 - 22603 Hamburg - Hamburg
}

\end{center}

\vspace{2cm}

\begin{abstract}
During the last year many important results
have been achieved in heavy flavour physics: New measurements of charm
and beauty production have been performed at HERA and the Tevatron.
A wealth of new spectroscopy data 
with several new, unexpected states in the charmonium and the $D\sub s$ 
systems has been collected and  $b\to d \gamma$ transitions have been
established. The oscillation frequency in the $B\sub s \bar B\sub s$
is now measured, and mixing in the $D^0 \bar D^0$ system has been
observed. Theoretical progress in the areas of open heavy flavour
production, quarkonium production and decays, and multiquark
spectroscopy has been presented at this workshop.
\end{abstract}

\vspace{1.5cm}

\noindent
{\footnotesize
To appear in the Proceedings of the XV International Workshop 
on
Deep-Inelastic Scattering and Related Subjectes, DIS 2007, April 16--20, 2007, Munich, Germany.
}

\end{titlepage}

\newpage

\section{Experimental Summary}

In this section we summarize the experimental results from the 
heavy flavour working group \cite{talk:list}. The presentations covered
a wide range of topics, from charm, beauty and charmonium production in $ep$ and $p
\bar p$ collisions, heavy ion results on charm suppression,
spectroscopy and rare decays, over oscillations in the 
$B\sub s \bar B\sub s$ and $D^0 \bar D^0$ systems to the outlook for
heavy flavour physics at future experiments at the LHC and the ILC.

\subsection{Charm and Beauty Production}

\subsubsection{Charm Production}

\begin{figure}[bth]
\centerline{\includegraphics[width=0.5\columnwidth,bbllx=0,bblly=70,bburx=450,bbury=709]{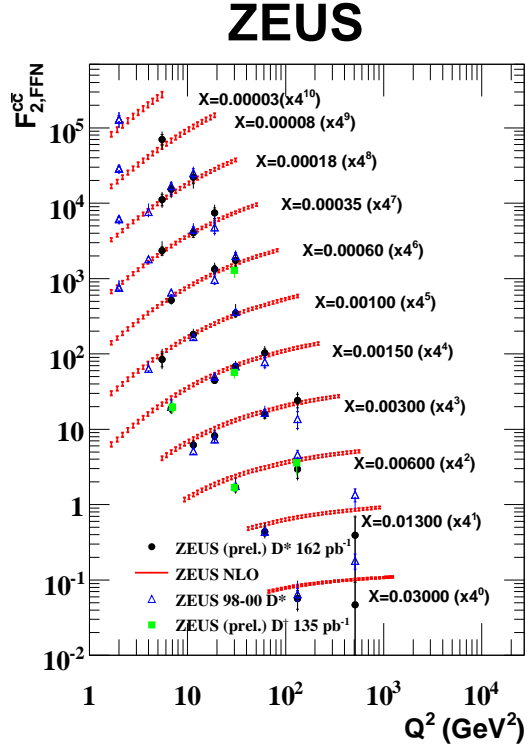}}
\caption{Compilation of $F_2\super{c}$ measurements from ZEUS.}
\label{fig:zeus-f2c}
\end{figure}

Charm Production in $ep$ collisions has been studied extensively over
the last years at HERA, and a wealth of data exist in photoproduction
(where the exchanged photon is quasi-real, with a virtuality $Q^2
\approx 0$) and deep-inelastic scattering (DIS).
The HERA collider experiments, H1 and ZEUS, have presented new preliminary 
results on
charm production from the HERA-II running period 
\cite{talk:stadie,talk:nicholass, talk:lipka}. 
The ZEUS Collaboration reported on two charm measurements in DIS with
HERA-II data (Fig.~\ref{fig:zeus-f2c}): One analysis  \cite{talk:stadie} uses $D^*$ mesons 
to identify charm production, and utilizes $162\,\pb^{-1}$ of new data
to achieve improved statistical accuracy compared to previous analyses.
The second  analysis \cite{talk:nicholass} uses $D^\pm$ mesons instead
and is one of the first measurements to utilize the new silicon strip 
Micro Vertex
Detector (MVD) of ZEUS. Based on $135\,\pb^{-1}$, this analysis achieves
similar accuracies as previous ZEUS measurements of the inclusive charm
cross section in DIS.

The ZEUS collaboration has recently also finalized two analyses of charm
production from older HERA-I data \cite{talk:stadie}.
One analysis \cite{Chekanov:2007mr} covers the region of very low
momentum transfer ($0.05 < Q^2 < 0.7\,\GeV^2$) at the transition
between photoproduction and DIS. 
The data provide a good test of perturbative QCD calculations, which are
available in NLO, and are well described by massive calculations in the
fixed flavour number scheme, 
as implemented in the HVQDIS program \cite{Harris:1997zq}.
In the second analysis \cite{Chekanov:2007ch}, the pseudoscalar states
$D^0$, $D^+$, and $D\sub s^+$ are reconstructed, rather than the vector
state $D^{*+}$,
which allows a measurement of the fragmentation fractions, which
 turn out to be compatible with those measured in $e^+e^-$
annihilation and in photoproduction \cite{talk:fang, Chekanov:2005mm}.
 The inclusive charm cross section derived from this
measurement is consistent with previous results.
Also a new, preliminary measurement of charm fragmentation was presented by
the ZEUS collaboration \cite{talk:fang}, which shows broad agreement
with other measurements from H1 \cite{talk:lipka}, and also with
measurements from $e^+e^-$ experiments. 

Another new measurement of $D^*$ production in DIS, based on
$222\,\pb^{-1}$ of HERA-II data,
was presented by the H1 collaboration
\cite{talk:lipka}. 
Here, differential and double differential visible cross
sections for $D^*$ production were measured and compared to 
the QCD calculations.
Overall, these quantities are well described
by the NLO predictions of HVQDIS; in fact, the data uncertainties are
in many cases smaller than the theory uncertainties from the variation
of the charm mass 
and the
variation of the renormalization and factorization scales.
However, in some quantities, most notably the 
$\eta\,(D^*)$ distribution, even in spite of the relatively large theory
uncertainties the NLO predictions deviate significantly from the trend
observed in data.

This confirms, with higher statistics, the observation made in a 
recently published analysis
from the H1 collaboration \cite{talk:schmidt, Aktas:2006py}.
However, that analysis goes one step further and investigates $D^*$
production in DIS in conjunction with jets. As heavy flavour production
is dominated by boson gluon fusion, a two-jet structure is expected for
most of the events. In the DIS analysis, one jet in addition to the
$D^*$ meson is required. Again, HVQDIS describes the data
satisfactorily, with the notable exception of the azimuthal angle
difference $\Delta \phi$ between the $D^*$ and the jet, 
a quantity which would be almost exactly equal to $180^\circ$ in leading
order, where the two charm jets must be back-to-back; hard gluon
emission, which enters only in NLO, leads to an enhancement
at lower values of $\Delta \phi$, which is underestimated by
HVQDIS, whereas Monte Carlo models such as CASCADE \cite{Jung:2001hx},
which include parton showers as approximation for higher-order gluon
radiation, work significantly better.
In another recent publication \cite{talk:schmidt, Aktas:2006ry}, the H1
collaboration has investigated charm photoproduction with two jets, with
very similar findings. 

The results show that charm measurements have
now reached a level of accuracy where more precise  QCD predictions
would be desirable. 
On one hand a full NLO Monte Carlo
with matched parton showers is needed, as opposed to HVQDIS, which
produces only parton four-vectors and has been augmented with relatively
simple independent fragmentation routines to make the extraction of
visible cross sections possible. A second way to reduce theory
uncertainties would be to reduce the variation of input parameters such
as the charm mass or fragmentation parameters by detailed comparisons
between MC predictions and data of (double) differential visible cross
sections in terms of relevant quantities such as $\pt\,(D^*)$ and
$\eta\,(D^*)$.
On the other hand, NNLO calculations would provide the most reliable way
to reduce theory uncertainties,  in particular for inclusive quantities
such as $F_2^{c \bar c}$. 

\subsubsection{Beauty Production}

Beauty production at HERA and the Tevatron has generated a lot of
interest in recent years, not least because at both colliders inclusive cross
sections for beauty production were observed that are considerably
higher than the expectation from perturbative QCD calculations in
next-to-leading order.

Achim Geiser gave an overview over the existing experimental results
from HERA, with additional glimpses at Tevatron and UA1 data \cite{talk:geiser}. 
He discussed in detail the issue of scale dependence of QCD
calculations, where 
he presented a survey of QCD calculations in different orders of 
$\alpha\sub s$ for a number of processes and argued that a good choice
for the factorisation and renormalisation scales $\mu\sub {r, f}$ 
would be a scale at
which higher order calculations give a result that coincides with the
calculations at lower order. He observed that this is achieved in most
of the cases at a scale between $1/4 \mu_0$ and $1 \mu_0$, where $\mu_0$
is the ``natural'' choice of scale such as $E\sub T^2$, $Q^2+m\sub Q^2$,
or $p\sub t^2+m\sub Q^2$. 
Based on this observation he argued that the factorisation and renormalisation 
scales should be varied in the range 
$1/4 \mu_0 < \mu\sub {r, f} < 1 \mu_0$, with a choice of 
$\mu\sub {r, f} = 1/2 \mu_0$ for the central value, rather than using
a variation of  $1/2 \mu_0 < \mu\sub {r, f} < 2 \mu_0$, as is customary
in most HERA publications. He also pointed out that this prescription
has already been adopted silently by the Tevatron experiments in recent
preliminary results. This proposal was received well by the audience, as
the discussion after the presentation showed.

New, preliminary results on beauty production were presented by the D0 and CDF
collaborations \cite{talk:reisert}.
D0 has performed a new measurement of muon tagged jet cross sections
based on $300\,\pb^{-1}$ of data that extends up to $p\sub t=420\,\GeV$.
Again, the data lie above the theory prediction.
In a new measurement of inclusive B cross sections, the CDF
collaboration utilizes the decays $B\to \ell D^0, D^{*+} X$ in addition
to the previously used channel
and $B\to J/\psi X$, with consistent results. These measurements are also
consistent with a fixed order calculation at next to leading log.

The first, preliminary HERA-II result on beauty production, based on
$39\,\pb^{-1}$ of data, was 
presented by the ZEUS collaboration \cite{talk:kahle}. The results are
somewhat higher, but still consistent with previous measurements from
H1 \cite{bib:h1f2cb}.

\subsection{Charmonium Production}

The inclusive production of charmonium and bottomium states in $ep$ and 
$p \bar p$ collisions remains an intersting testing ground for perturbative 
QCD, as was shown in a review by Katja Kr\"uger \cite{talk:krueger}.
The working horse for these investigations are 
 $J/\psi$ mesons, where the largest data sets are
available.
While leading order calculations in the colour singlet model
\cite{bib:csm-lo} fail to describe the production rates of
$J/\psi$ mesons at the Tevatron \cite{bib:tevatron-jpsi},
NRQCD models \cite{Bodwin:1994jh} models have been successfully applied
there. 
In these models the production of $c\bar c$ states with different colour
and spin/parity configuration is calculated perturbatively, and the
transition to bound states is described by non-perturbative 
long distance matrix elements (LDME), which have to be determined from
data. This allows the production of $J/\psi$ mesons by gluon 
splitting into a $c\bar c$ pair, 
followed by a transition to a
 $J/\psi$ meson, which is predicted to be transversely polarized.
 The observation of such a polarization is therefore considered the
``smoking gun'' for (large) colour octet (CO) contributions.

However, new data from CDF that were
presented at this conference \cite{talk:reisert} show a clear evidence
for longitudinal $J/\psi$ polarization, 
which is in contrast to NRQCD
expectations. New data on $\psi'$ polarisation are still 
not precise enough for firm conclusions.

A second test of NRQCD predictions is the measurement of the production
rates of $\chi\sub{c2}$ versus $\chi\sub{c1}$ states, which in NRQCD
models is expected to follow the spin counting prediction of $5/3$.
New measurements from CDF \cite{Abulencia:2007br, talk:reisert}
give a significantly smaller result for the ratio of prompt 
$\chi\sub c$ production:
$
  \sigma(\chi\sub{c2}) / \sigma(\chi\sub{c1}) = 0.70 \pm 0.04(stat.)
  \pm 0.04 (syst.) \pm 0.06 (BF).
$

After the initial success of the NRQCD model indications for the
necessity of large CO contributions where also searched for in inelastic
$J/\psi$ production at HERA.
The H1 collaboration has presented new, preliminary data from 
HERA-II on inelastic $J/\psi$ production in deep inelastic $ep$
scattering \cite{talk:steder}. 
Production rates were measured double differentially in the
transverse momentum $p\sub t$ and the momentum fraction $z$ of the $J/\psi$
meson, and compared to Monte Carlo predictions by programs which
implement the colour singlet model in leading order.
These models describe the shape of the measurements quite well, whereas
the CO contributions in NRQCD models tend to have different $p\sub t$
and $z$ shapes than the CS contributions. Therefore, the new H1 data
do not show any clear need for large CO contributions to inelastic  
$J/\psi$ production. In the discussion of the results the need was
stressed for NLO calculations of this process, in the CS as well as in
NRQCD models. NLO calculations are currently only available 
in the CS model for photoproduction of $J/\psi$ mesons \cite{Kramer:1995nb}, 
where they
describe the data from HERA \cite{bib:hera-jpsi-photoprod} quite well.

A different issue in charmonium production is adressed by the new data
from HERA-B \cite{talk:zurnedden}: HERA-B has collected large samples of
$J/\psi$ and $\psi'$ mesons decaying to $\mu^+\mu^-$ and $e^+e^-$ pairs,
which made it possible to investigate the dependence of charmonium
production on the atomic weight $A$ in a new range of Feynman-$x\sub F$,
extending the range covered by the experiments down to $x\sub F =
-0.35$, which is a region where theoretical models
make widely different predictions.

\subsection{Heavy Ion Results}

The new results from heavy ion experiments at RHIC have been reviewed in
this conference by William Zajc \cite{talk:zajc}. 
Therefore we only briefly highlight some of the new results presented in the heavy
flavour session.

New results on heavy quark production in Au$+$Au collisions were presented by the
PHENIX \cite{talk:hornback} and the STAR \cite{talk:mischke} collaborations.
Due to their larger mass and the dead cone effect, charm quarks are expected to
lose energy at a smaller rate than light quarks.
The observation of supression of electrons from charm decays in 
central Au$+$Au collisions 
\cite{talk:hornback, Adare:2006nq, talk:mischke, Abelev:2006db} 
therefore comes as a surprise. 
It appears that charm quarks participate in the flow of the opaque 
hadronic medium just as much as light quarks, which means that they thermalize more
quickly than expected in many theoretical models.

The PHENIX collaboration has also presented new measurements of $J/\psi$ suppression
in Au$+$Au collisions \cite{talk:atomssa, Adare:2006ns}. 
These measurements, which are performed in two rapidity ranges $|
y|<0.35$ and $1.2<y<2.2$, show a significant suppression of $J/\psi$ production in 
Au$+$Au collisions, which is stronger at large rapidities than at central rapidity
values. The explanation of this rapidity dependence is a real challenge to
theoretical models.

\subsection{Spectroscopy and Rare Decays}

The large data sets from the B factory experiments BaBar and Belle, from
CLEO-c and from the Tevatron experiments CDF and D0 have lead to a
renewed interest in the spectroscopy of charm and bottom hadrons with
beautyful results, and have opened the possibility to identify extremely
rare decays.

\subsubsection{The Charmonium System}

In the charmonium system, finally all expected charmonium states below the
D-meson threshold have now been firmly identified. One of last two missing
states, the $\eta\sub{c}'$, has now seen by BaBar, CLEO-c, and
Belle \cite{talk:seth, bib:etacprime}, at an average mass of
$m(\eta\sub{c}') = 3638 \pm 4\,\MeV$.
This allows a comparison of the hyper fine splitting of the $1S$ and
$2S$ states in the charmonium system, which are
$\Delta m\sub{hf}\,(1S) = 117\pm 1\,MeV$ and 
$\Delta m\sub{hf}\,(2S) = 48\pm 4\,MeV$; the large difference of these 
values poses a challenge to theory \cite{talk:seth}.

The long elusive $h\sub{c}$ state has observed by CLEO-c 
\cite{talk:seth, Rosner:2005ry} in the decay chain $\psi(2S) \to \pi^0
h\sub{c}$, $h\sub{c} \to \gamma \eta\sub c$,
with a mass of $m(h\sub{c}) = 3524.4\pm0.6\pm0.4 \MeV$,
which corresponds to a hyper fine splitting between the
$h\sub{c}$ and the center of gravity of the $\chi\sub{c 0,1,2}$ states
of $m\sub{hf} (1P) = +1.0 \pm 0.6\pm 0.4 \MeV$, consistent with the expected
value of zero. 
Meanwhile, CLEO-c has increased its $\psi(2S)$ sample
eightfold, which yield the promise of further, improved results.

In addition to these expected charmonium states, recent years have seen
 the discovery of several unexpected charmonium-like resonances: 
 
New results were obtained on the $X(3872)$, which is already considered 
firmly
established by the PDG \cite{Yao:2006px},
and on the $Y(3940)$ and $Y(4260)$.
 Results were also presented on two additional
states, the $X(3940)$ and the $Z(3930)$.

For the $X(3872)$, one  explanation that has been put forward is
the interpretation as a $D^0 \bar{D^{0*}}$ molecule.
CLEO has performed a new measurement of the $D^0$ mass \cite{talk:seth}:
$m(D^0)= 1864.847 \pm 0.150 (stat.) \pm 0.095 (syst.) \MeV$ 
\cite{Cawlfield:2007dw}. The total uncertainty of this measurement of
$0.178\,\MeV$ is a factor of 2.2 better than the uncertainty of
$0.4\,\MeV$ of the 2006 world average \cite{Yao:2006px}. 
Combinig this result with the 
PDG06 value of the $X(3872)$ mass of $3871.2\pm0.5\,\MeV$ results in a
very small binding energy of $E\sub{b} = 0.6 \pm 0.6 \MeV$ for a 
$D^0 \bar{D^{0*}}$ molecule \cite{talk:seth}.
Meanwhile, Belle and BaBar have also found indications for the
$X(3872)$ in the decays $B\to \bar D^0 D^0 \pi^0 K$ (Belle) and
$B \to \bar D^0 D^{0*}  K$ (BaBar) \cite{talk:poireau}.
In these channels, the observed mass values for the $X(3872)$ of
$m=3875.4 \pm 0.7 ^{+1.2}_{-2.0}\,\MeV$ (Belle) \cite{Gokhroo:2006bt}
and $m= 3875.6 \pm 0.7 ^{+1.4}_{-1.5}\,\MeV$ (BaBar)
are $2.5\,\sigma$ higher than the current world average.
Belle also concludes that the quantum numbers 
$J^{PC}=1^{++}$ are favoured if the observed enhancement is indeed the 
$X(3872)$. 
All in all, the interpretation of this state remains unclear.

New data were also presented on the $Y(4260)$ \cite{talk:poireau},
which was first observed in initial state radiation events at BaBar. 
BaBar sets a limit \cite{Aubert:2006mi} of $BR(Y(4260) \to D \bar D)/BR(Y(4260) \to 
J/\psi \, \pi^+ \pi^-) < 7.6$, which is 
a further indication that the $Y(4260)$ is
not a conventional charmonium state.
CLEO has also confirmed the $Y(4260)$ \cite{talk:seth,Coan:2006rv},
and finds \cite{He:2006kg} its mass to be 
$m= 4284^{+17}_{-16}(stat)\pm 4 (syst.) \MeV$,
in poor agreement 
with the original BaBar measurement \cite{Aubert:2005rm}
of $m= 4259 \pm 8 ^{+2}_{-6}\,\MeV$.
The latest Belle \cite {talk:poireau,Abe:2006hf} result  
$m= 4295 \pm 10 ^{+10}_{-3}\,\MeV$ 
is consistent with the CLEO measurement.

The discovery of new, unexpected charmonium-like states has also
triggered new investigations of $R=\sigma (e^+e^-\to hadrons)/\sigma
(e^+e^-\to \mu^+\mu^-)$ at Belle and CLEO \cite{talk:pakhlova,
Coan:2006rv, He:2006kg, Abe:2006hf}. These scans show a marked dip
around the $Y(4260)$. Belle, BaBar and CLEO have also looked 
\cite{talk:pakhlova,Aubert:2006mi, Abe:2006fj,Mendez:2007ey}
into the
more exclusive channels $e^+e^- \to D^{(*)} D^{(*)}$. 
A weak signal for the $Y(4260)$ is seen in the $D \bar D^*$ channel, 
no signal in the
$DD$ channel, and a dip, similar to the one observed in the inclusive
$R$ measurement, is observed in the $D^* D^*$ channel around the
$Y(4260)$ resonance.

Meanwhile, BaBar observes yet another state 
\cite{talk:poireau, Aubert:2006ge} in initial state production of
$e^+ e^- \to \psi(2S) \pi^+ \pi^-$ at a mass of
$m=4324\pm24\,\MeV$ and a width of $\Gamma = 172\pm33\,\MeV$, which is
incompatible with the $Y(4260)$ or other known states such as the $\psi(4415)$.

Three more states observed by Belle \cite{talk:poireau, bib:xyz3940},
termed $X(3940)$, $Y(3940)$, and $Z(3930)$, may have an interpretation
as conventional charmonium states, namely the
 $\eta\sub{c}(3S) [3\,^1 S_0]$, the $\chi'\sub{c1} [ 2\,^3P_1]$ and
 the $\chi'\sub{c2} [2\,^3 P_2]$.

\subsubsection{Charmed, Strange Mesons}
 
In the sector of charmed, strange mesons new measurements were presented
by BaBar and Belle \cite{talk:poireau}.
In addition to new measurements of the properties of the
$D\sub{s0}^*(2317)$ and $D\sub{s1}(2460)$, another new state, termed
$D\sub{sJ}^*(2860)$, has been identified by BaBar
\cite{Aubert:2006mh}
with a mass 
$m= 2856.6 \pm 1.5 \pm 5.0 \MeV$
and a spin parity assignment
$J^P=0^+, 1^-, 2^+, ...$. In addition, a hint for another state
with $m= 2688 \pm 4 \pm 3 \MeV$ 
has also been observed.
Furthermore, Belle, in a Dalitz analysis of the decay
$B^+ \to \bar D^0 D^0 K^+$, sees indications for a state
$D\sub{sJ} (2700)$ with $m = 2715 \pm 11 ^{+11}_{-14}\,\MeV$,
with quantum numbers
$J^P=1^-$ favoured.
The theoretical interpretation of these states is not yet clear.
While the $D\sub{s0}^*(2317)$ and $D\sub{s1}(2460)$ states can be
explained as $0^+$ and $1^+$ P-wave $c \bar s$ states, their masses
are substantially lower than expected from potential models. The
interpretation of the other states is still less clear.

\subsubsection{Charmed Baryons}

Coming to charmed baryons, BaBar and Belle have reported discoveries of
several new states \cite{talk:pakhlova}.
Both observe the new state $\Lambda\sub{c}(2940)^+$ in the channels
$D^0 p$ (BaBar) and $\Lambda\sub{c}^+ \pi^+ \pi^-$ (Belle) \cite{bib:charmbaryons}.
The Belle discovery of the new charmed, strange baryons
$\Xi\sub{c}(2980)^+$ and $\Xi\sub{c}(3077)^+$ has been confirmed by BaBar
\cite{bib:charmstrangebaryons}, and Belle sees also some evidence for the
isospin partners $\Xi\sub{c}(2980)^0$ and $\Xi\sub{c}(3077)^0$.
BaBar has also reported the first observation of the $\Omega\sub{c}^*$,
an excited $css$ state \cite{Aubert:2006je}. 

\subsubsection{Bottom Mesons}

In the B meson sector, the progress comes from the Tevatron experiments
CDF and D0 \cite{talk:heck}. The $B\sub{c}^+$ has already been observed 
by both
experiments, now CDF reports the first direct observation of the
$B\sub{c}^+$  in the exclusive decay channel 
$B\sub{c}^+ \to J/\psi \, \pi^+$, which allows a very precise mass
measurement with the preliminary result of 
$m(B\sub{c}^+) = 6276.5 \pm 4.0 \pm 2.7\,\MeV$, while the uncertainty on
the old world average was $400\,\MeV$.
The $\eta\sub b$ is last spin singlet ground state in the bottomium
sector that has not been unambiguously discovered yet. 
CDF has performed a new search for the decay 
$\eta\sub b \to J/\psi \, J/\psi$, without indications for a signal.
Both, D0 and CDF, report first direct observations on orbitally excited
B mesons with $L=1$ by looking for decays $B^{**} \to B^{(*)+}\pi^-$. 
Both see clear evidence for 
$B_1$ and $B_2$ states. 
With a similar analysis of the channels  $B\sub{s}^{**} \to B^{(*)}+ K^-$ 
both experiments observe also the $B\sub{s2}^*$ state, 
in addition CDF reports 
evidence for the $B\sub{s1}$ state.

\subsubsection{Bottom Baryons}

In the bottom baryon sector, where up to now the $\Lambda\sub b^0$ is
the only well established particle, CDF performed a blind analysis of 
the decay channel $\Lambda\sub b^0 \pi^\pm$; after unblinding, four
resonances were found with significances greater than $5\,\sigma$, which
constitute the first direct observation of $\Sigma\sub b^{(*)}$ baryons.
The resonances are identified as the $\Sigma\sub b^\pm$ and the $\Sigma\sub b^{*\pm}$.

\subsubsection{Rare Decays}

In the field of rare decays, the large data sets of altogether more than 1 billion
$B \bar B$ events obtained at the B 
factory experiments allow more and more precise measurements of $b \to s
\gamma$ decay modes \cite{talk:eschrich}, which are interesting because
in the Standard Model they are forbidden at tree level and proceed
through loop diagrams. In extensions of the Standard Model, additional
particles contribute to the loops, which may lead to observable
deviations of the transition rates from the expected SM values.
While the $b \to s \gamma$ decay channels are investigated with higher
and higher precision, the large statistics now allows even the measurement
of $b \to d \gamma$ transitions, which were first observed by Belle in
2005 \cite{Abe:2005rj}.
The latest compilation from the Heavy Flavour Averaging Group \cite{Barberio:2007cr}
contains 20 measured $b \to s/d \gamma$ decay channels, with branching
fractions as small as $4.6\cdot 10^{-7}$ for $B^0 \to \omega \gamma$
and precisions down to $6.5\,\%$ for $B^+ \to K^+ \pi^+ \pi^- \gamma$.
The 2006 average for $b \to s \gamma$ decays \cite{Barberio:2006bi} is
now $BR(B \to X\sub s \gamma) = (3.55 \pm 0.26) \cdot 10^{-4}$, which is on
the high side of recent NNLO predictions \cite{talk:eschrich,bib:bsgammannlo}. 
The latest measurements from BaBar and Belle of $B \to \rho/\omega
\gamma$ \cite{bib:brhoomegagamma} are particularly impressive, measuring
branching fractions around $10^{-6}$, in some cases with more than
$5\,\sigma$ significance.
From a combination of the $b \to d/s \gamma$ measurements from Belle and
BaBar, a constraint of $|V\sub{td}/V\sub{ts}| =
0.202^{+0.017}_{-0.016} (exp.) \pm 0.015(theor.)$ has been derived.

A very difficult, but interesting decay channel is $B^+ \to \tau^+ \bar \nu$
\cite{talk:eschrich,bib:btotau}, which has been observed by Belle at
$3.5\,\sigma$ significance with a branching ratio of 
$BR\,(B^+ \to \tau^+ \bar \nu) = (1.79
^{+0.56}_{-0.49}\,^{+0.39}_{-0.46}) \cdot 10^{-4}$,
while BaBar measures 
$(0.88^{+0.68}_{-0.67}\pm 0.11) \cdot 10^{-4}$, or, translated into a 
$90\,\%$~CL limit, $BR\,(B^+ \to \tau^+ \bar \nu) < 1.8\cdot 10^{-4}$.
The combination of both results yields 
$(1.31 \pm 0.48) \cdot 10^{-4}$, corresponding to a $2.5\,\sigma$
evidence. 
This result can be used to derive limits on the $H^\pm$ mass in SUSY
models.

After the observation of $B\sub s$ oscillations, the search for the
decays $B\sub {s,d} \to \mu^+ \mu^-$ 
might be considered the next race in B physics.
The Standard Model expectations for these decays are extremely small,
and probably out of reach for current experiments:
$BR(B\sub s \to \mu^+ \mu^-) =(3.42 \pm 0.54) \cdot 10^{-9}$ and
$BR(B\sub d \to \mu^+ \mu^-) =(1.0 \pm 0.14) \cdot 10^{-10}$.
Again, these decays can only proceed through loop diagrams in the SM,
and contributions from new particles may increase the rate by orders of
magnitude \cite{talk:corcoran}.
Both Tevatron experiments have searched for the $B\sub s \to \mu^+ \mu^-$ decay and have
reported new, preliminary results based on new Run-II data.
The limits are 
$BR(B\sub s \to \mu^+ \mu^-) < 10 \cdot 10^{-8} (9.3 \cdot 10^{-8})$ 
from CDF (D0); these limits
correspond to branching ratios that are 29 (27) times larger than the
SM expectation.
CDF has also reported a preliminary limit for $B\sub{d}$ decays: 
$BR(B\sub d \to \mu^+ \mu^-) < 2.3 \cdot 10^{-8}$, which is 230 times
greater than the SM expectation.

\subsection{Mixing and Oscillations}

While oscillations have long been established and thoroughly
investigated in the $K^0 \bar K^0$ and $B^0 \bar B^0$ systems,
until recently the frequency of the $B\sub s \bar B\sub s$
oscillations had not been measured, and no firm signal for mixing in the
$D^0 \bar D^0$ system had been observed. 
During the last year, both these gaps
in our knowledge of neutral meson mixing have been filled.

The $B\sub s$ oscillation frequency has now been measured by the
Tevatron experiments \cite{talk:kehoe}.
After the first report on a double sided limit for the $B\sub s$
oscillation frequency of $17 < \Delta m\sub s < 21\,\ps^{-1}$ at
$90\,\%$~CL from the D0 collaboration \cite{Abazov:2006dm},
CDF has for the first time observed a nonzero oscillation amplitude with
more than $3\,\sigma$ significance \cite{Abulencia:2006mq}, at a frequency 
$\Delta m\sub s = 17.77 \pm 0.10 \pm 0.07\,\ps^{-1}$,
consistent with the
D0 result.

Meanwhile, the D0 collaboration has gone several steps further.

Based on a sample of $1.1\,\fb^{-1}$ of decays 
$B\sub s \to J/\psi \, \phi$, D0 has made a new measurement 
\cite{Abazov:2007tx}
of the difference between the lifetimes of the long and short lived
$B\sub s$ eigenstates.
Both states can decay to the $J/\psi \, \phi$
final state, and by fitting the decay angle distributions their
respective contributions to the sample at different eigentimes can be
determined, with the result
$\Delta \Gamma \sub s = 0.17 \pm 0.09 \pm 0.02\,\ps^{-1}$.
For the first time, the D0 collaboration has used this data also to
extract the CP violating phase $\phi\sub s$, which is the relative phase
of the off-diagonal elements of the mass and decay matrices in the
$B\sub s \bar B \sub s$ basis, from this data:
$\phi\sub s = -0.79 \pm 0.56 (stat.) ^{+0.14}_{-0.01} (syst.)$.
The SM prediction for $\phi\sub s$ is very small, namely
$\phi\sub s = (4.2 \pm 1.4)\cdot 10^{-3}$  \cite{Lenz:2006hd}.

D0 has also reported on a new measurement \cite{Abazov:2007rb} of the branching ratio
$BR\,(B\sub s \to D\sub{s}^{(*)} D\sub{s}^{(*)}) = 0.039
^{+0.019}_{-0.017} (stat.)\,^{+0.016}_{-0.017} (syst.)$, which is lower
than the only pre-existing measurement from ALEPH.
This branching ratio is linked theoretically to the 
width difference
\begin{figure}[bth]
\centerline{\includegraphics[width=0.45\columnwidth]{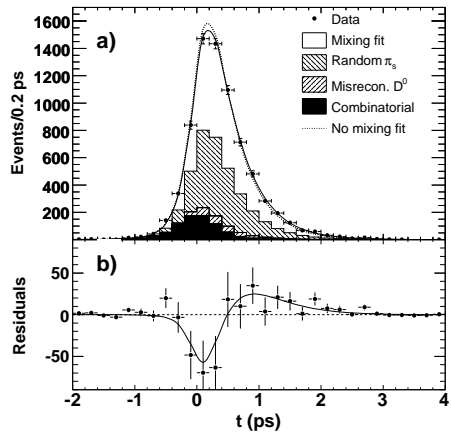}}
\caption{Proper time distribution of the $D^0 \to K^+ \pi^-$ (wrong-sign WS) decays from
BaBar}\label{fig:babar-d0-ws}
\end{figure}
$\Delta \Gamma \sub s \super{CP}$ between the CP-even and odd $B\sub s$
eigenstates. The resulting constraint is
$\Delta \Gamma \sub s \super{CP} / \Gamma \sub s = 0.079
^{+0.038}_{-0.035} (stat.)\,^{+0.031}_{-0.030} (syst.)$. 

While the difficulty in $B\sub s $ mixing lies in the fact that the
mixing is almost perfect
because the oscillations
occur much faster than the decay, the situation is reversed in the $D^0
\bar D^0$ system. Here, the Standard Model prediction for the mixing
parameters $x=\Delta m/\Gamma$ and $y=\Delta \Gamma / 2\Gamma$ are very
small \cite{talk:pakhlova}, of the order $10^{-6...-2}$, which means
that the $D^0$ decays much faster than one oscillation period lasts. Again,
additional particles in the loop may increase the mixing, which could
indicate new physics. In particular, new physics processes such as
Flavour-Changing Neutral Currents, SUSY particles etc. would increase
$x$ (i.e., the oscillation frequency), but not $y$ (the lifetime
difference).

Until march of this year, the no-mixing case had been disfavoured with a
significance of $2.1\,(2.3)\,\sigma$ by Belle (BaBar) analyses
\cite{bib:oldd0mixing}. Then both collaborations published 
two papers titled ``Evidence for $D^0 \bar D^0$ mixing'' side by side
in Physics Review Letters
\cite{bib:d0mixing}, 
both reporting more than $3\,\sigma$ evidence for 
$D^0 \bar D^0$ oscillations.

The Belle analysis is based on $540\,\fb^{-1}$ of data and measures the
difference in apparent lifetime for the CP even decays 
$D^0 \to K^+K^-, \pi^+\pi^-$ 
to the lifetime of the decay 
$D^0 \to K^-\pi^+$.
The data show indications for mixing with $3.2\,\sigma$ significance.
In addition, the Belle collaboration has performed a Dalitz analysis of
the decay $D^0 \to K\sub S^0 \pi^-\pi^+$, which provides the most
sensitive result on $x$ to date: $x=(0.80 \pm 0.29 \pm 0.17)\,\%$
and disfavours the no-mixing solution with $2.7\,\sigma$ significance
\cite{talk:pakhlova, Abe:2007rd}.

The BaBar analysis uses $384\,\fb^{-1}$ of data and analyses the decays
$D^0 \to K^- \pi^+, K^+ \pi^-$. The dominant decay mode is the  
right-sign (RS) decay mode to $K^- \pi^+$, while the wrong-sign (WS) decay
mode to $K^+ \pi^-$ may occur as doubly Cabibbo-suppressed (DCS) decay
or through mixing. These two mechanisms are separated by the analysis of
the time dependence of the decay, see Fig~\ref{fig:babar-d0-ws}.
From this data, BaBar can exclude the no-mixing hypothesis with
$3.9\,\sigma$.

\subsection{Future Experiments}

Looking ahead into the future, first beams at the LHC are now expected
in early 2008. Heavy flavour physics will be of interest at the
experiments ATLAS, CMS \cite{talk:krasznahorkay} and LHCb
\cite{talk:santovetti}.

In ATLAS and CMS, the expected rates for beauty production are high: at
a luminosity of $\lumi=10^{33}\,\percmsqs$, around $10^5$ $b \bar b$
pairs will be produced per second, which should allow high precision
measurements despite the difficult environment. The big challenge here
is to set up a reasonably efficient trigger, because bottom quarks are
predominantly produced at relatively low transverse momenta, while the
detectors are optimized for high-$\pt$ discovery physics.

In the initial, ``low'' luminosity phase with expected luminosities 
around $\lumi=10^{33}\,\percmsqs$, the ATLAS experiment envisages to use
single particle triggers at Level 1, which will be refined by searches
for more complex signatures in the High Level Trigger (HLT) 
\cite{talk:krasznahorkay}. Channels that have been studied include 
e.g. $B\sub s \to D\sub s \pi$ with subsequent decays $D\sub s \to \phi
\pi$, states involving a electromagnetic signature such as $J/\psi
\to ee, K^* \gamma, \phi \gamma$, and final states with two muons.
In the high luminosity mode at $\lumi=10^{34}\,\percmsqs$, Level 1 will
predominantly use the dimuon channel to identify $J/\psi$ or $B \sub s$
decays to $\mu\mu$.
CMS will also use muon triggers at Level 1, and plans to identify decay
vertices from heavy flavour decays in the High Level trigger.
A number of topics have been studied by both collaborations, such as the
measurement of inclusive beauty cross sections (CMS), the measurement of
$\sin\,2\beta$ in the golden mode $B^0 \to J/\psi \, K^0\sub S$ (ATLAS), analysis
of $B\sub s \to J/\psi \, \phi$ decays (ATLAS and CMS), and searches for rare
decays such as $\Lambda\sub b \to \Lambda \mu \mu$, $B^0 \to K^* \mu
\mu$ or $B\sub s \to \phi \mu \mu$ (ATLAS).
The ATLAS collaboration has also studied the prospects to measure $B\sub s$ oscillations.
They conclude that they could measure $\Delta m\sub s$  with $5\,\sigma$
significance at the current CDF value with the data from one year, i.e.
$10\,\fb^{-1}$.

CMS has also studied the prospects to measure the $B\sub c$ mass from
decays to $J/\psi \, \pi$; the expected resolution from $1\,\fb^{-1}$ of
data is $22\,\MeV$ statistical and $15\,\MeV$ systematical uncertainty,
which can be compared to the preliminary CDF result reported in this conference
 \cite{talk:heck}, based on $1.1\,\fb^{-1}$, with statistical and
 systematic errors of $4.0$ and $2.7\,\MeV$, respectively. 
 
The LHCb experiment \cite{talk:santovetti}, is an experiment dedicated
to the study of B physics at LHC. The detector is a single arm
spectrometer, optimized to detect b hadrons at pseudo rapidities $1.9 <
\eta < 4.9$, and will run at an interaction zone with a relatively low
luminosity to reduce backgrounds, where one year of running at nominal
luminosity will provide $2\,\fb^{-1}$ of data.
The LHCb collaboration has studied the prospects to measure the angles
of the unitarity triangle; for $\sin 2 \beta$ they hope to achieve a
statistical precision of $0.02$ from one year of data taking in the
golden channel $B^0 \to J/\psi \, K^0\sub S$. For the least well determined
angle $\gamma$, the most promising method seems to be the analysis of
decays $B^{0,+} \to D^0 K^{0,+}$ with $D^0$ decays to $K\pi, K 3\pi,
\pi\pi$ and $KK$, with a comparison of the rates of Cabibbo Favoured 
and Doubly
Cabibbo Suppressed decays. The estimated sensitivities are in the range 
$\sigma(\gamma) \approx 5^\circ - 15^\circ$, depending on the actual
value of $\gamma$.

An  interesting benchmark is provided by the 
sensitivity of $B\sub{d, s} \to \mu \mu$, which has been studied by
ATLAS, CMS \cite{talk:krasznahorkay}, and LHCb \cite{talk:santovetti}.
Recall that currently CDF and D0 have reported upper limits of 
$100 \cdot 10^{-9}$
for the branching ratio \cite{talk:corcoran} based on
samples of $2\,\fb^{-1}$ (D0) and $0.78\,\fb^{-1}$ (CDF); 
these limits lie about 30 times above
the SM expectation of $3.4 \cdot 10^{-9}$.
A recent ATLAS study concludes that with an integrated luminosity of
$30\,\fb^{-1}$, corresponding to about three year's running in the low
luminosity mode,
an upper limit of 
$6.6 \cdot 10^{-9}$ at $90\,\%$ CL is achieveable, which is about a 
factor of two above the SM expectation.
CMS concludes that it could provide a limit of 
$14 \cdot 10^{-9}$ (four times the SM expectation) with 
$10\,\fb^{-1}$, i.e. one year's worth of data. 
LHCb on the other hand, which will operate at lower luminosities, claims
a sensitivity that would allow a $3\,\sigma$ observation of the decay 
$B\sub s \to \mu \mu$ at the Standard Model branching ratio of 
$3.4 \cdot 10^{-9}$ with $2\,\fb^{-1}$, corresponding to one
year's worth of data \cite{talk:santovetti} at nominal luminosity.

Looking even farther into the future, Tim Greenshaw reported on the
applications of heavy flavour at the planned Innternational Linear
Collider (ILC) \cite{talk:greenshaw}, an $e^+e^-$ collider with a
centre-of-mass energy between $500$ and $1000\,\GeV$. 
The detector designs for this machine forsee a very good flavour
separation power, which calls for vertex detectors with spectacular
performance. Several groups are currently investigating various
technologies that could fulfill the requirements for the inner pixel
detector of an ILC detector. 
One important application of the heavy flavour identification
capabilities of such a detector would be the measurement of the various
branching fractions of a Higgs boson. Even if only one neutral  Higgs
boson were found at the LHC, this would allow to check wether these
fractions are consistent with the predictions from the Standard Model,
where only one Higgs doublet is assumed and therefore all branching
fractions depend only on the mass of the single physical  Higgs boson,
or whether for instance up- and down-type fermions have different
couplings to the Higgs, as expected for instance in Supersymmetric models,
which have at least two Higgs doublets.

\subsection{Conclusions for the Experimental Part}

Heavy flavour production at HERA and Tevatron remains an interesting
field of research, both from the experimental and the theory side. 
The experimental results become more and more precise,
which makes it interesting to include them in the
determination of parton densities, provided that the theoretical
obstacles can be overcome and uncertainties arising from charm mass and
fragmentation functions are treated consistently between experiment and
theory. The beauty production data have taught us that the calculation
of production cross sections do not always become more accurate for 
heavier quarks if there is sufficient phase space for
QCD dynamics, which is something to keep in mind for LHC, where also the
top will fall into this category.

During the last years we have seen a veritable renaissance of hadron
spectroscopy. As the particle data book fills up, we see more and more
results that indicate how incomplete our understanding of hadron
structure still is, as illustrated by unexpected differences in
hyperfine splittings in the charmonium sector \cite{talk:seth}, by
unexpected states and masses in the $c\bar s$ system \cite{talk:poireau}, and
by the appearance of charmonium-like resonances for which we have no
ready explanation \cite{talk:seth,talk:poireau,talk:pakhlova}. 
While
resonances that do not fit into the conventional $q \bar q$ picture have
been known for a long time in the light meson sector, we now learn that 
also in the heavy quark sector there is more than is written in our
philosophy. On the other hand, the investigation of rare decays such as
$B^+ \to \tau \nu$ or $B \to \mu\mu$ and other rare processes such as
$D^0 \bar D^0$ mixing, where we wait for results that would
point to physics beyond the Standard Model, 
has once again failed to turn up anything unexpected.

\section{Theory}

In this section, we summarize the six theory contributions to the
heavy-flavor working group, emphasizing computations that have been
performed during the years 2006 and 2007 and have therefore not yet been
presented at previous DIS workshops \cite{talk:klasen}. We start with new
perturbative results for inclusive final states, i.e.\ heavy-quark structure
functions and their relation to parton densities, and then move on to less
inclusive final states, in particular quarkonium production and decay and
heavy-quark spectroscopy. For the latter we emphasize new results from
lattice QCD and QCD sum rules. Note that a series of new calculations on
open heavy-quark production at various colliders in the general-mass
variable-flavor number scheme (GM-VFNS) has been presented at this workshop
in an introductory plenary talk by G.\ Kramer \cite{kramer}.

The heavy-quark coefficient functions for deep-inelastic lepton-hadron
scattering (DIS) in the kinematic regime $Q^2 \gg m^2$ have been
calculated more than ten years ago in next-to-leading order (NLO) of QCD by
M.\ Buza et al., using operator product expansion techniques
\cite{Buza:1995ie}. Here $Q^2$ and $m^2$ stand for the masses squared of the
virtual photon and heavy quark respectively. The analytical results had been
expressed in terms of 48 independent functions and had been used to check
earlier, general calculations, which were, however, only accessible via
large computer programs.
J.\ Bl\"umlein has now presented a re-calculation of the 
${\cal O}(\alpha_s
^2)$ massive operator matrix elements for the twist-2 operators, which
contribute to the heavy flavor Wilson coefficients in unpolarized DIS in the
region $Q^2\gg m^2$, using light-cone expansion techniques and confirming
the above-cited calculation \cite{Bierenbaum:2007qe}. The application of the
integration-by-parts method and harmonic sums in Mellin space allowed for a
significant compactification of the results, which can now be expressed in
terms of the basis $\{S_1,~S_2,~S_3,~S_{-2},~S_{-3}\}$ and $S_{-2,1}$, i.e.\
of only two independent functions.

\begin{figure}[thb]
 \centerline{\includegraphics[width=0.45\columnwidth]{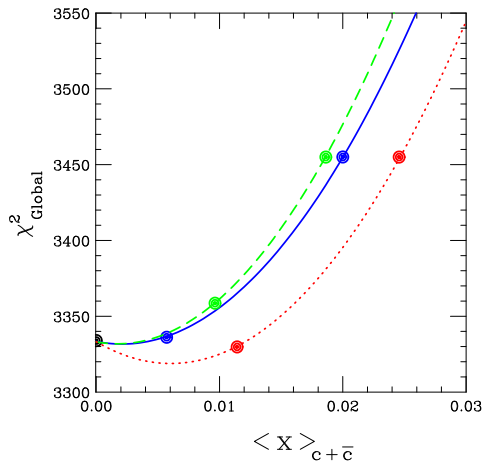}}
 \caption{Goodness-of-fit vs.\ momentum fraction of intrinsic charm at the
 starting scale $\mu=1.3$ GeV for the light-cone (solid curve), meson-cloud
 (dashed curve), and sea-like model (dotted curve).}\label{hfl:1}
\end{figure}

While the proton is just a simple $|uud\rangle$ Fock state in the quark
model, the possibility of an intrinsic-charm, i.e.\ a $|uudc\bar{c}\rangle$,
component has repeatedly been put forward in the context of light-cone
\cite{Brodsky:1980pb} or meson-cloud models \cite{Navarra:1995rq,Melnitchouk:1997ig}. 
W.K.\ Tung et al.\ have performed global fits of
parton density functions (PDFs), assuming that the charm-density is not only
generated radiatively at $\mu=m_c$ and then evolved to $Q$, but allowing for
the possibility of light-cone, meson-cloud, or sea-quark like intrinsic
charm density \cite{Pumplin:2007wg}. The quality of each fit is measured by
a global $\chi^2$, shown in Fig.\ \ref{hfl:1} as a function of the momentum
fraction $\langle x\rangle_{c+\bar{c}}$ carried 
 by the charm quark at the starting scale $\mu=m_c=1.3$ GeV. For $\langle x\rangle
_{c+\bar{c}}\leq0.01$, the quality of the fit varies very little, i.e.\ the
global analysis of hard-scattering data 
provides no evidence either for or
against intrinsic charm. Above this point, all three curves in Fig.\
\ref{hfl:1} rise steeply with $\langle x\rangle _{c+\bar{c}}$, so that
global fits do place useful upper bounds on intrinsic charm. There is
no data set that is particularly sensitive to intrinsic charm,
but the global QCD analysis rules out the possibility of an intrinsic charm
component much larger than $0.02$ in momentum fraction. A variation of the
charm quark mass $m_c$ shows that the data prefer lower masses around 1.3
GeV with respect to higher masses of 1.5 GeV. The difference in $\chi^2$ is
in this case almost entirely due to the charm contribution to the proton
structure function $F^2$, which has been precisely measured at HERA.

Turning to less inclusive quantities, the production and decay of heavy
quark-antiquark bound states (quarkonia ${\cal Q}$) is still far from
understood. While the effective field theory of non-relativistic QCD (NRQCD)
has long been believed to be phenomenologically successful and is still
believed to be theoretically more consistent than the color-singlet model
(CSM), recent Tevatron Run II data on the polarization of large-$p_T$ $J/
\Psi$- and $\Upsilon$-mesons do not support the idea that their production
is dominated by color-octet fragmentation processes as predicted in NRQCD.
Since higher-order QCD corrections are largely unknown (except for
color-singlet photoproduction \cite{Kramer:1994zi}, color-singlet and
color-octet production in photon-photon collisions \cite{Klasen:2004tz,Klasen:2004az} 
and a very recent calculation for direct color-singlet
hadroproduction \cite{Campbell:2007ws}), other theoretical frameworks
continue to be investigated. One example is the $k_T$-factorization
formalism, where the production cross section
\begin{eqnarray}
 &&{\rm d}\sigma(p\bar{p}\to{\cal Q}+X)~=~
 \int{{\rm d}x_1\over x_1}\int{\rm d}k_{1T}^2\int{{\rm d}\phi_1\over2\pi}
 \Phi(x_1,k_{1T}^2,\mu^2)\\&&\hspace*{17mm}\times
 \int{{\rm d}x_2\over x_2}\int{\rm d}k_{2T}^2\int{{\rm d}\phi_2\over2\pi}
 \Phi(x_2,k_{2T}^2,\mu^2)\
 {\rm d}\hat{\sigma}(RR\to{\cal Q}+X)\nonumber
\end{eqnarray}
is computed by a double convolution of unintegrated parton densities $\Phi$
and partonic cross section $\hat{\sigma}$ over longitudinal momentum
fractions $x_{1,2}$ and intrinsic transverse momenta $k_{1T,2T}$ of the
reggeized partons $R$ in the (anti-)proton $p$ ($\bar{p}$). This method has
been applied by Kniehl et al.\ to charmonium production at different
colliders \cite{Kniehl:2006sk} and now, as V.A.\ Saleev reported, also to
bottomonium hadroproduction at the Tevatron \cite{Kniehl:2006vm}. The
long-distance operator matrix elements (OMEs) of NRQCD have been refitted to
the $p_T$-spectra of prompt $\Upsilon$ mesons at the Tevatron. Since the
intrinsic $k_T$ leads to a harder $p_T$-spectrum already for the
color-singlet contributions, the color-octet OMEs turn out to
be considerably smaller and in many cases even consistent with zero, as
their values are not sufficiently constrained by the data. In addition,
the results depend strongly on the assumed unintegrated parton densities,
which are poorly known and only constrained to agree with the integrated
ones,
\begin{eqnarray}
 xf(x_{1,2},\mu^2)&=&\int_0^{\mu^2}{\rm d}k_{1T,2T}^2\,\Phi(x_{1,2},
 k_{1T,2T}^2,\mu^2).
\end{eqnarray}
As can be seen from Tab.\ \ref{hfl:2}, the quality of the fit, measured
\begin{table}[h]
 \begin{center}
 \caption{\label{hfl:2}OMEs of the $\Upsilon(1S,2S,3S)$ and $\chi_{b0}(1P,
 2P)$ mesons from fits to CDF data from Runs I and II in the
 Regge-kinematics approach using unintegrated gluon distributions by
 J.\ Bl\"umlein (JB), Jung and Salam (JS), and Kimber, Martin and Ryskin
 (KMR).}\vspace*{2mm}
 \begin{tabular}{|c|ccc|}
 \hline
 NME & Fit JB & Fit JS & Fit KMR \\
 \hline
$\langle{\cal O}^{\Upsilon(1S)}[^3S_1^{(1)}]\rangle/$GeV$^3$ &
 $10.9\pm 1.6$ &  $10.9\pm 1.6$ & $10.9\pm 1.6$ \\
$\langle{\cal O}^{\Upsilon(1S)}[^3S_1^{(8)}]\rangle/$GeV$^3$ &
$(5.3\pm 0.5)\times 10^{-3}$ & $(0.0\pm 1.8)\times 10^{-4}$ &
$(0.0\pm 3.1)\times 10^{-3}$ \\
$\langle{\cal O}^{\Upsilon(1S)}[^1S_0^{(8)}]\rangle/$GeV$^3$ &
$(0.0\pm 4.7)\times 10^{-4}$ & $(0.0\pm 5.2)\times
10^{-5}$ &
$(0.0\pm 4.3)\times 10^{-3}$ \\
$\langle{\cal O}^{\Upsilon(1S)}[^3P_0^{(8)}]\rangle/$GeV$^5$ &
$(0.0\pm 1.3)\times 10^{-3}$ & $(0.0\pm 1.6)\times
10^{-4}$ &
$(9.5\pm 2.0)\times 10^{-2}$ \\
$M_{5}^{\Upsilon(1S)}/$GeV$^3$ &
$(0.0\pm 7.6)\times 10^{-4}$ & $(0.0\pm 8.7)\times 10^{-5}$ &
$(2.1\pm 0.9)\times 10^{-2}$\\
$\langle{\cal O}^{\chi_{b0}(1P)}[^3P_0^{(1)}]\rangle/$GeV$^5$ &
$2.4\pm 0.4$ & $2.4\pm 0.4$ & $2.4\pm 0.4$ \\
$\langle{\cal O}^{\chi_{b0}(1P)}[^3S_1^{(8)}]\rangle/$GeV$^3$ &
$(0.0\pm 2.1)\times 10^{-3}$ & $(0.0\pm 8.4)\times 10^{-5}$ &
$(0.0\pm 1.4)\times 10^{-3}$ \\
$\langle{\cal O}^{\Upsilon(2S)}[^3S_1^{(1)}]\rangle/$GeV$^3$ &
$4.5\pm 0.7$ & $4.5\pm 0.7$ & $4.5\pm 0.7$ \\
$\langle{\cal O}^{\Upsilon(2S)}[^3S_1^{(8)}]\rangle/$GeV$^3$ &
$(0.0\pm 5.9)\times 10^{-3}$ & $(0.0\pm 4.1)\times 10^{-4}$ &
$(3.3\pm 0.8)\times 10^{-2}$ \\
$\langle{\cal O}^{\Upsilon(2S)}[^1S_0^{(8)}]\rangle/$GeV$^3$ &
$(0.0\pm 9.2)\times 10^{-4}$ & $(0.0\pm 8.3)\times
10^{-5}$ &
$(0.0\pm 3.7)\times 10^{-3}$ \\
$\langle{\cal O}^{\Upsilon(2S)}[^3P_0^{(8)}]\rangle/$GeV$^5$ &
$(0.0\pm 2.6)\times 10^{-3}$ & $(0.0\pm 2.8)\times
10^{-4}$ &
$(0.0\pm 1.6)\times 10^{-2}$ \\
$M_{5}^{\Upsilon(2S)}/$GeV$^3$ &
$(0.0\pm 1.5)\times 10^{-3}$ & $(0.0\pm 1.4)\times 10^{-4}$ &
$(0.0\pm 7.2)\times 10^{-3}$ \\
$\langle{\cal O}^{\chi_{b0}(2P)}[^3P_0^{(1)}]\rangle/$GeV$^5$ &
$2.6\pm 0.5$ & $2.6\pm 0.5$ & $2.6\pm 0.5$ \\
$\langle{\cal O}^{\chi_{b0}(2P)}[^3S_1^{(8)}]\rangle/$GeV$^3$ &
$(1.1\pm 0.4)\times 10^{-2}$ & $(0.0\pm 2.8)\times 10^{-4}$ &
$(0.0\pm 5.7)\times 10^{-3}$ \\
$\langle{\cal O}^{\Upsilon(3S)}[^3S_1^{(1)}]\rangle/$GeV$^3$ &
$4.3\pm 0.9$ & $4.3\pm 0.9$ & $4.3\pm 0.9$ \\
$\langle{\cal O}^{\Upsilon(3S)}[^3S_1^{(8)}]\rangle/$GeV$^3$ &
$(1.4\pm 0.3)\times 10^{-2}$ & $(5.9\pm 4.2)\times 10^{-3}$ &
$(1.1\pm 0.4)\times 10^{-2}$ \\
$\langle{\cal O}^{\Upsilon(3S)}[^1S_0^{(8)}]\rangle/$GeV$^3$ &
$(0.0\pm 2.6)\times 10^{-3}$ & $(0.0\pm 8.1)\times
10^{-4}$ &
$(0.0\pm 2.7)\times 10^{-3}$ \\
$\langle{\cal O}^{\Upsilon(3S)}[^3P_0^{(8)}]\rangle/$GeV$^5$ &
$(2.4\pm 0.8)\times 10^{-2}$ & $(3.4\pm 4.2)\times
10^{-3}$ &
$(5.2\pm 1.1)\times 10^{-2}$ \\
$M_{5}^{\Upsilon(3S)}/$GeV$^3$ &
$(5.2\pm 4.4)\times 10^{-3}$ & $(7.4\pm 10.2)\times 10^{-4}$ &
$(1.1\pm 0.5)\times 10^{-2}$\\
$\langle{\cal O}^{\chi_{b0}(3P)}[^3P_0^{(1)}]\rangle/$GeV$^5$ &
$2.7\pm 0.7$ & $2.7\pm 0.7$ & $2.7\pm 0.7$ \\
$\chi^2/\mathrm{d.o.f.}$ & $2.9$ & $27$ & $0.5$ \\
\hline
 \end{tabular}
 \end{center}\vspace*{-3mm}
\end{table}
by $\chi^2$ per degree of freedom, varies widely with the PDFs from 0.5 to
27.

Quarkonium decay OMEs can be computed from first principles using lattice
QCD, so that their determination becomes independent of fits to experimental
data or potential model assumptions. A.\ Hart reported about a recent result
that matches the electromagnetic vector-annihilation current in lattice
NRQCD to the one in continuum QCD, which should allow for a prediction of
the leptonic decay widths of $S$-wave bottomonia with ten percent accuracy
and of the ratio of the 2S and 1S $\Upsilon$-states with one percent
accuracy \cite{Hart:2006ij}. Numerical results should be available soon.

Charmonium ($\eta_c$ and $\chi_{c0}$) decays into two photons have been
calculated for the first time by Dudek and Edwards, albeit only in the
quenched approximation, using relativistic valence quarks and a perturbative
expansion of the photon-quark coupling, which allows to replace the photon
by a superposition of QCD states \cite{Dudek:2006ut}. While the width
obtained for the $\chi_{c0}$ ($2.41\pm0.58\pm0.72\pm0.48$ keV) is in rather
good agreement with the experimental value ($2.84\pm0.40$ keV), the one for
the $\eta_c$ ($2.65\pm0.26\pm0.80\pm0.53$ keV) is smaller by a factor of
three ($7.14\pm2.49$ keV). This might be due to an incorrect running of the
strong coupling constant and a depleted wave function at the origin in this
calculation. In addition, the discretization error might not be reliable,
as only one lattice spacing has been used.

Lansberg and Pham have computed the two-photon widths of ground-state and
radially excited $\eta_c$ \cite{Lansberg:2006dw} and $\eta_b$ mesons
\cite{Lansberg:2006sy} with an effective Lagrangian in the static
approximation, taking into account binding energy effects for the radial
excitations. Using heavy-quark spin symmetry, they assume equality of
the $f_{\eta_c}$ $(f_{\eta_{c'}})$ and $f_{J/\Psi}$ ($f_{\Psi'}$) decay
constants and relate the two-photon width of the $\eta_c$ to the leptonic
decay width of the $J/\Psi$. While good agreement with the experimental
value is found for the $\eta_c$ ground state (7.5-10 keV), their result
for $\eta_c'$ (3.5-4.5 keV) is three times larger than the CLEO measurement
($1.3\pm0.6$ keV). This may be due to the fact that $f_{\eta_c'}$ is
not equal, but three times smaller than $f_{\Psi'}$, according to a recent
lattice calculation in the quenched approximation by Dudek and Edwards
\cite{Dudek:2006ej}.

Radiative decays of charmonia to light mesons have recently been computed in
perturbative QCD \cite{Gao:2006bc}, albeit keeping only the color-singlet
wave function contribution and assuming the light mesons to be also
described by non-relativistic color-singlet $q\bar{q}$ bound states with
finite constituent quark masses. These have then been used to regularize the
one-loop diagrams. The numerical results are given in Tab.\ \ref{hfl:3}.
Unfortunately a systematic study on the
\begin{table}[h]
 \begin{center}
 \caption{\label{hfl:3}Proposed theoretical and measured experimental values
 for ${\cal B}(J/\psi\to f_J\gamma)$.}\vspace*{2mm}
 \begin{tabular}{ |c|ccccc|}
 \hline &$f_0(980)$&$f_1(1285)$&$f_2(1270)$&$f_1'(1420)$&$f_2'(1525)$ \\\hline
 ${\cal B}_{th}\times10^4$&$1.6$&$7.0$&$8.7$&$1.8$&$2.0$\\
 ${\cal B}_{ex}\times10^4$&-&$6.1\pm0.8$&$13.8\pm1.4$
 &$7.9\pm1.3$&$4.5^{+0.7}_{-0.4}$\\ \hline
 \end{tabular}
 \end{center}\vspace*{-3mm}
\end{table}
theoretical uncertainties with the chosen masses and scales and with respect
to possible color-octet contributions has not been performed. The calculated
$J/\psi$-decay branching ratios to the $P$-wave mesons $f_2(1270)$ and
$f_1(1285)$ fit the data well, while that of $f_0(980)$ (if treated as
an $s\bar s$ meson) is predicted to be $1.6\times 10^{-4}$, which implies
that $f_0(1710)$ can not be the $s\bar s$ or $(u\bar u+d\bar d)/\sqrt{2}$
meson. Decays of $P$-wave charmonia $\chi_{cJ}\to \rho(\omega, \phi)\gamma$
($J=0,1,2$) have also been studied. The branching ratio of $\chi_{c1}\to
\rho\gamma$ is predicted to be $1.4\times 10^{-5}$, which may be tested by
CLEO-c and BESIII with future experiments.

A similar calculation has been performed for the radiative decays of
bottomonia into charmonia and light mesons, taking into account in addition
the sometimes significant QED contributions \cite{Gao:2007fv}. While the
results for radiative decays into truly non-relativistic charmonia are
likely to be more reliable than those for radiative decays into (in reality
relativistic) light mesons, these decays remain to be observed. On the other
hand, the calculated branching ratios for $\Upsilon\to f_2(1270)\gamma$
($6.3\times10^{-5}$) and $\Upsilon\to f_2'(1525)\gamma$ ($2.0\times10^{-5}$)
are only barely consistent with recent CLEO data [$(1.0\pm0.1)\times10^{-4}$
and $(3.7\pm1.2)\times10^{-5}$], suggesting that the theoretical approach
may still need improvement.

Finally, J.M.\ Richard reviewed the current status of light and heavy
multiquark spectroscopy \cite{Richard:2006md}. While experimental evidence
for the $uudd\bar{s}$ pentaquark $\theta^+$ seems to vanish, a single-charm
baryon with a mass of 2940 MeV and a width of approximately 17 MeV has
recently been confirmed by BaBar, and a double-charm baryon with a mass of
3520 MeV may have been observed by the Selex collaboration. Potential models
have since long predicted the existence of baryons with at least one heavy
quark, assuming them to be diquark-quark systems that can be described with
hyperspherical coordinates. In the meson sector, the state $X(3940)$ may be
a $1^{-+}$ candidate for a $ccg$ hybrid state, to be described in the
Born-Oppenheimer approximation with classical constituent gluons. A second
recently confirmed state is the $X(3872)$, which QCD sum rules predict to be
a $1^{++}$ $c\bar{c}q\bar{q}$ bound state, even if the theoretical mass
found is slightly larger ($3925\pm127$ MeV) \cite{Matheus:2006xi}.
 
\section{Acknowledgments}

We thank all the speakers of the heavy-flavor working group for their
contributions and the local committee for the perfect organization of
the DIS 2007 workshop.


\begin{footnotesize}

\end{footnotesize}
  

\end{document}